# NanoCarb part 2: Performance assessment for total column $CO_2$ monitoring from a nano-satellite

Silvère Gousset[*a], Laurence Croizé[b], Etienne Le Coarer[a], Yann Ferrec[b], and Laure Brooker[c] for the SCARBO consortium[d]

[a]Institut de Planétologie et d'Astrophysique de Grenoble, Université Grenoble-Alpes, 38058 Grenoble, France;
[b]ONERA/DOTA, BP 80100, chemin de la Hunière, 91123 Palaiseau, France;
[c]Airbus Defence and Space, 31, rue des Cosmonautes, 31402 Toulouse, France;
[d]http://scarbo-h2020.eu/

**ABSTRACT**

NanoCarb is an innovative Fourier Transform imaging spectrometer dedicated to the measurement of $CO_2$ and $CH_4$. Both its unusual optical principle and sampling strategy allows to reach a compact design, ideal for small satellite constellation as investigated by the European project SCARBO. The NanoCarb performance assessment as well as a proof of concept are required in this framework. A strategy of design is developed to optimize the performances and reach the sensitivity target of the space mission, demonstrating the potential of the concept without drastic complexity gain. A preliminary bias mitigation in the retrieval strategy is presented concerning water for $CO_2$ measurement, illustrating the efficiency and the flexibility of the NanoCarb partial interferogram sampling technic. The presented design reaches a random error sub-ppm for $CO_2$ and sub-10ppb for $CH_4$, considering a 128 to 192 km swath, respectively, for 2 or 3 km of resolution at ground. A full mitigation of the water bias is performed on $CO_2$ band thanks to partial interferograms.

**Keywords:** Air pollution monitoring, Anthropogenic GHG emissions, Fabry-Perot, Fourier transform spectroscopy, Optical instrumentation, Passive remote sensing

## 1. INTRODUCTION

One of the main issue of the study of climate changes due to human activities is to reduce the uncertainty over emission of the main GreenHouse Gas (GHG), $CO_2$ and $CH_4$ [1]. A first challenge for dedicated space missions is to monitor the atmospheric total column with both a drastically improved statistical error and absolute accuracy <sub-ppm (part per million) for the concentration of $CO_2$, and of a few parts per billion (ppb) for $CH_4$. The second challenge is to reach a daily revisit and an Earth global coverage in a few days.

Currently, MicroCarb [2] and CarbonSat [3] are the two future missions dedicated to the passive monitoring of GHG from space. Nevertheless, these two missions cannot reach a sufficient spatial and temporal coverage. In this framework, the Horizon 2020 *Space CARBon Observatory* (SCARBO) [4] aims at assessing the feasibility of a low-cost constellation. The project relies on small satellites in a sun-synchronous orbit to monitor $CO_2$ and $CH_4$ emissions, complementing the low-revisit high-performance satellites.

The core miniature sensor of this constellation is the NanoCarb concept, a Fourier Transform imaging spectrometer [5]. Both the use of a low finesse Fabry-Perot array and a partial interferometric sampling strategy allow to achieve a large swath at high spectral resolution as well as an optimal use of the available pixels for an high sensitivity in a snapshot acquisition mode.

For the SCARBO project we first aim at maturating the technological readiness level of the NanoCarb concept, from lab development to prototype integration and inflight proof of concept. In parallel, we also assess NanoCarb performances for the space mission itself. In this paper, we present the preliminary performance assessment for the space mission.

---

[*] silvere.gousset@univ-grenoble-alpes.fr

We expect to first explain what is the design strategy of this uncommon concept. Then, how we manage bias mitigation with the NanoCarb specificities in a retrieval model. Finally, we investigate the expected performances of the NanoCarb concept for the SCARBO project, mainly focused on $CO_2$.

## 2. NANOCARB CONCEPT PRINCIPLE

In this section, we describe the principle of the NanoCarb spectrometer: its optical principle, then its sampling strategy, and finally the different data products of NanoCarb in the SCARBO project.

### 2.1 Optical principle

NanoCarb is a static Fourier Transform imaging spectrometer. We can distinguish two optical sub-systems: the front optics and the interferometric core, presented in the scheme on Figure 1.

The front optics consist in an afocal system with a field stop, allowing to manage independently of the interferometric core the spatial sampling over the Focal Plan Array (FPA), as well as the size of the total imaged field of view (fov). An interference filter selects the spectral band. Four independent spectral bands have been currently identified for SCARBO as reported in Tab. 1. In the scope of this paper we consider only the B2 and B3 bands, assuming a clear sky and a perfect measurement of the atmospheric pressure thanks to the B1 band.

Table 1 Current NanoCarb bands

| Band | Region | Measurement |
|---|---|---|
| B1 – $O_2$ | 760 nm | [$O_2$], aerosols |
| B2 – $CO_2$ | 1.6 μm | [$CO_2$] |
| B3 – $CH_4$ | 1.66 μm | [$CH_4$] |
| B4 – strong $CO_2$ | 2.06 μm | aerosols |

The interferometric core is formed by the association of an interferometer array with a micro-lens (μlens) array. Each interferometer has a specific thickness. The FPA is placed in the focal plane of the μlens array, to obtain a set of thumbnail each one associated to one thickness of the interferometric plate. In this configuration, the image formed in each thumbnail is a replication of the same fov, modulated in intensity by the associated interferometer.

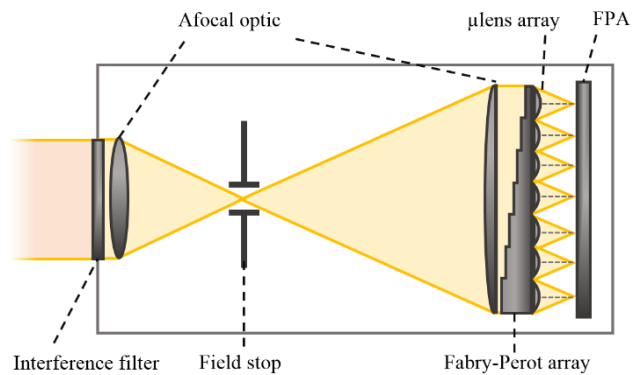

Figure 1 Optical principle of NanoCarb: a front afocal optic with a field stop is located before the core of NanoCarb, consisting in the interferometric plate and the lens array. Then, images are formed on the FPA to obtain a set of thumbnails, each one associated to one thickness of the interferometric plate. The interference filter selects the band. This optical system can be replicated with a different FP array and interference filter to simultaneously measure several spectral bands or interferometric signatures, with the same FPA.

Each interferometer is a low finesse Fabry-Perot (FP), which modulates the focal plane intensity with rings in the associated thumbnail, at a given Optical Path Difference (OPD). This configuration is optimal to manage both compactness of the device and large fov, as well as sensitivity as explained with more details in [6] of this conference.

The device can be replicated, each one filling a part of the same FPA, in order to monitor synchronously several spectral bands.

This concept is compact (length ~15-20 cm), fully static, and stable with a real capability to thermally regulate the full device. In the following, we will explain the partial interferometric sampling strategy of the NanoCarb concept, enabling snapshot acquisitions at high spectral resolution.

**2.2 Partial sampling of the interferogram**

The strategy we adopt to reach an high spectral resolution while keeping a large swath in snapshot mode is to target only the useful information in the Fourier space. Thus, the NanoCarb spectrometer acquires only partial interferograms.

The acquisition of partial interferograms was already developed in the 1970s by Kyle for temperature measurement in $CO_2$ lines [7], then by Fortunato who applied this method to the measurement of $SO_2$ concentration [8]. More recently, some studies can be found for the SIFTI instrument [9], and for IASI data processing [10], showing both a better geophysical bias mitigation like surface temperature and an improvement of the Signal-to-Noise Ratio (SNR).

In the case of NanoCarb, the spectral band is purposely selected to optimize the sensitivity over targeted regions of the interferogram, leading to an optical filtering of the useless information (interferants) both in the spectral and in the Fourier domain. This principle is well suited for $CO_2$ and $CH_4$, thanks to a periodic spectroscopic profile over specific spectral bands, for example on the 1.6 µm spectral band, or on the 1.66 µm band for $CH_4$ as visible on Figure 2. In these conditions, the signature of these two species in the Fourier domain is well concentrated as visible in Figure 3.

Hence, the aim of the partial interferogram sampling technic is:

- To obtain a measure of the X-specie concentration ([X]), as independently of the interferants as possible (for example water).
- To obtain a measure of the interferants where appropriate, as independently as possible. The main atmospheric interferants are the aerosols, the water vapor, the temperature profile and the pressure at ground.

As an illustration, Figure 3 shows some partial derivative interferograms (Jacobians), for a variation of concentration of $CO_2$ or $CH_4$, and of water, respectively. The red dashed lines highlight regions where the sensitivity to $CO_2$ or $CH_4$ is optimal, but biases by water, while the blue one spot regions where the measurement of water is optimal with respect to $CO_2$ or $CH_4$.

Figure 4 shows how with a controlled distribution of thickness over the FP array we can sample the interferogram at targeted OPD, potentially through disjoint interferometric regions. For a given region, the thicknesses are chosen to achieve a continuous λ/4-sampling. First, this allows to obtain a good sampling of the interferogram on a single acquisition, and thus enables snapshot acquisition mode; secondly it is more convenient for fringe contrast estimation based on generalized ABCD technics [11].

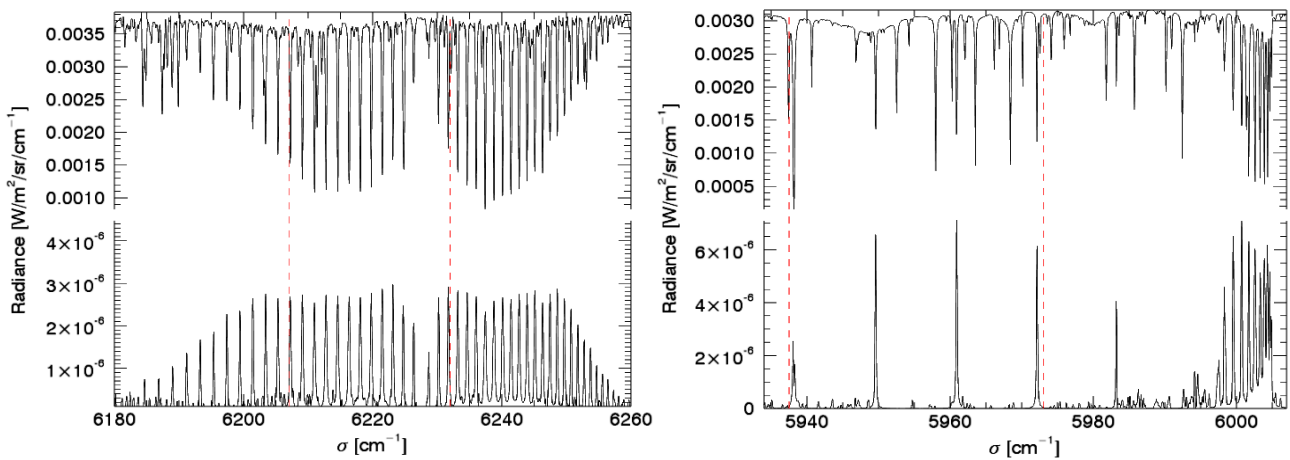

Figure 2 Simulated spectral radiance (top curve) and partial derivative radiance (bottom curve). Left: on 1.6 µm $CO_2$ band, with a variation of 1 ppm of $CO_2$ over the total column for the derivative radiance. Right: on 1.66 µm $CH_4$ band, with a variation of 10 ppb of CH$_4$. The spectral regions between the red dashed lines allow to optimize the sensitivity of NanoCarb and are targeted by the relevant interferometric filter.

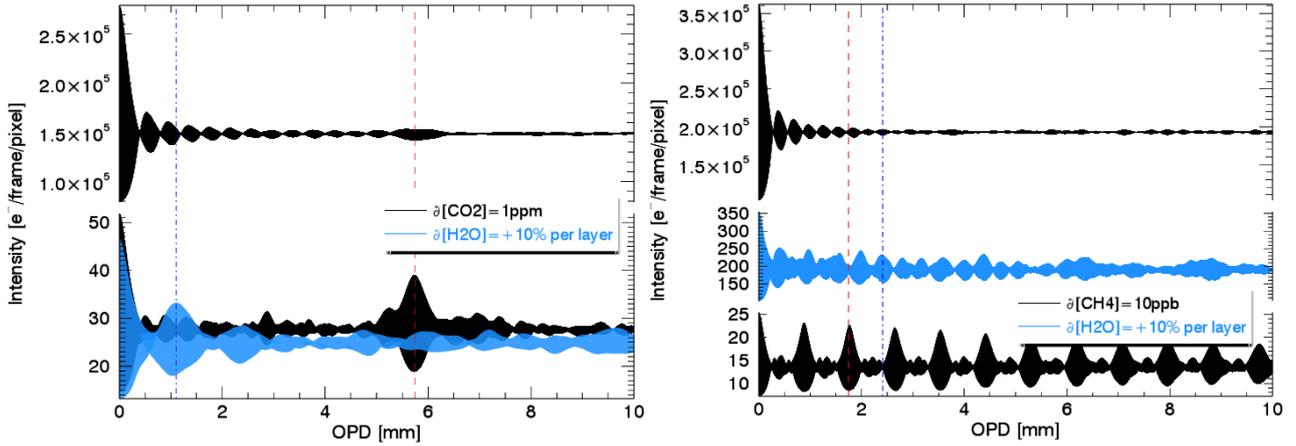

Figure 3 Interferogram (top curve) and partial derivative interferograms (bottom curve). Left: for the CO$_2$ band, with a variation of 1 ppm of CO$_2$ for the black-plotted derivative. Right: for the CH$_4$ band, with a variation of 10 ppb of CH$_4$ for the black-plotted derivative. In blue, the partial derivative for a variation of 10% of the total column of water for each band. The red dashed lines show the OPD where the specie sensitivity is optimal. The blue dotted-dashed lines show a maximum of sensitivity to water. Albedo=0.2, solar angle=20°, ifov=2 km.

To conclude, we can notice that the coherent flux in the red-spotted regions of the interferograms of Figure 3, $F_X$, is directly linked to the absorbed light power in the atmosphere, and thus to the atmospheric concentration respectively in CO$_2$ or CH$_4$. Moreover, we assume that:

$$\partial F_X \propto \partial [X] \qquad (1)$$

meaning that the absorbed light power by the specie X varies linearly with [X], that is well verified if we consider the unsaturated CO$_2$ and CH$_4$ lines in Figure 2. Hence, a measurement of [X] can be achieved by a retrieval in the Fourier space from an instrumental measurement of the fringe contrast in the targeted region.

### 2.3 NanoCarb data products

**The L0 data product** of NanoCarb is the snapshot focal plane intensity acquired by the detector. For demonstration purpose, a noiseless simulation can be seen in Figure 4-left, in the unrealistic case of a spatially and spectrally uniform scene.

The colored points spot a single elementary field of view (ifov) imaged in all the thumbnails. The extraction of the intensity on a single exposure for this ifov allows to retrieve the associated partial interferogram (Figure 4-right), assuming a lab calibration of the corresponding OPD (see [12] in this conference). This snapshot interferogram is the **L1a data product** of NanoCarb.

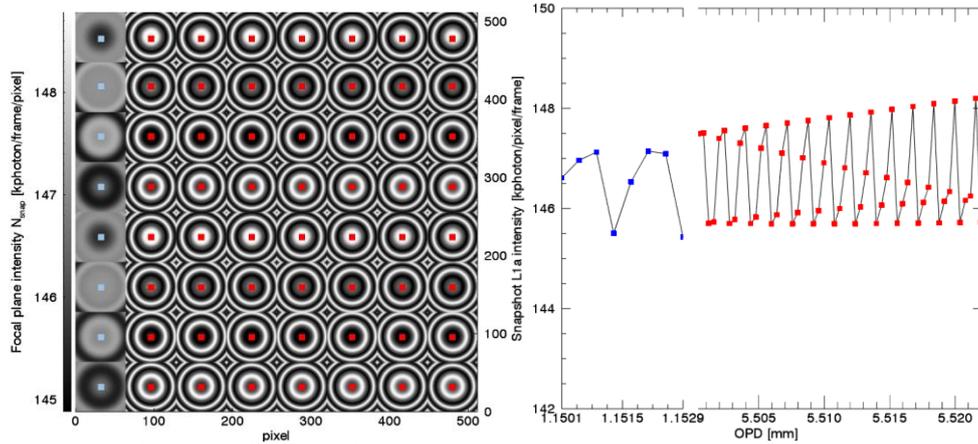

Figure 4 Left: simulated focal plane intensity of NanoCarb in the 1.6 μm $CO_2$ band, equivalent to L0 data product. Right: extracted snapshot interferogram for the spotted ifov, equivalent to L1a data product. In this configuration, we target two disjoint interferometric regions. The left thumbnail column is dedicated to the sampling of particularly water-sensitive fringes, while the rest of the thumbnails are dedicated to the $CO_2$.

The presented illustration Figure 5 allows to understand how to derive the **L1b data product** from a co-registration of the snapshot L1a data while the ifov shifts across the thumbnail. Figure 6 presents the obtained interferogram for a single ifov, after co-registration of acquired snapshots every time the ifov shifts by one pixel. We will demonstrate in this paper how this feature is useful to reach the targeted sensitivity for concentration measurement.

Finally, the **L2 data Product** of NanoCarb is the $CO_2$ and $CH_4$ concentration of the total atmospheric column, estimated from L1b.

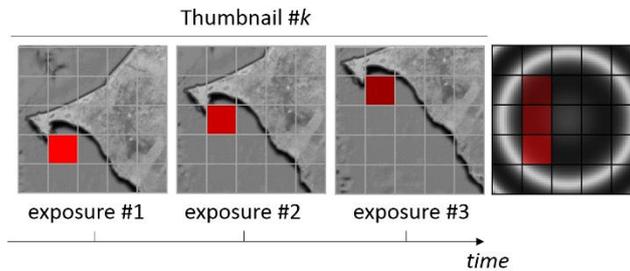

Figure 5 Ifov shifting across the thumbnail as a function of time. The right-image illustrates how the intensity is modulated over each exposure with the Fabry-Perot rings crossing. In that way, we can measure several interferometric states for ifov.

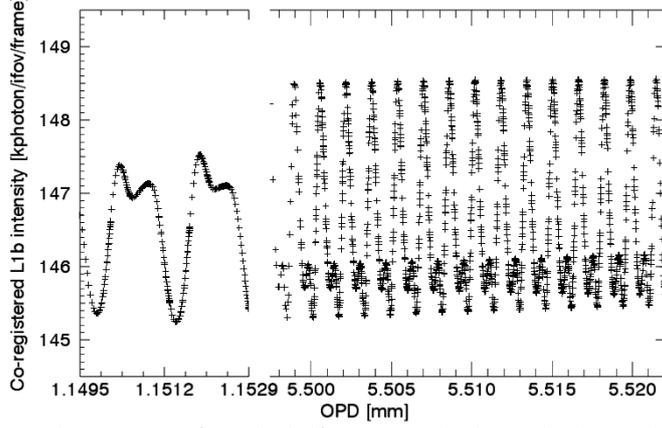

Figure 6 Co-registration of snapshot exposures for a single ifov. A snapshot is acquired every time the ifov shifts by one pixel, while the ifov is imaged in the total fov. The obtained interferogram is the L1b data product of NanoCarb for this single ifov.

## 3. RADIOMETRIC PERFORMANCES

We propose here a solution to estimate the sensitivity of NanoCarb as a function of the instrumental parameters. Thus, we expect to demonstrate the radiometric performances of NanoCarb and its potential for the SCARBO mission, expressed in terms of statistical error over the total column of $CO_2$ or $CH_4$ measurement. This study is based on a direct approach, overpassing noise propagation as well as a large amount of biases we preliminary investigate with a backward model in the next section. We limit our study to the SNR over the acquired signal, assuming a linear evolution of the absorbed power with the concentration. This last point is well suited in the unsaturated 1.6 µm and 1.66 µm band we consider here. We first introduce analytical expressions of the sensitivity, then we derive some performances with radiometric considerations.

### 3.1 NanoCarb analytical sensitivity

We explained in the previous section that our capability to measure a variation of concentration with NanoCarb relies on the accuracy over the estimation of the coherent flux in the targeted region of the partial interferogram, and thus, over the fringes visibility measurement. A solution to derive an expression of the radiometric sensitivity of NanoCarb is to compare the required sensitivity for a targeted concentration variation, to the effective SNR over the fringe visibility measurement.

The required sensitivity $S_{\partial[X]}$ to measure a given concentration variation $\partial[X]$ of the specie $X$ can be expressed as the ratio of variation of the local coherent flux $F_X$ in the targeted interferometric region:

$$S_{\partial[X]} = \frac{\partial F_X}{\partial[X]} / F_X \qquad (2)$$

Replacing $V_X = F_X / \bar{N}_{ph}$ the mean fringes contrast in the targeted Fourier region and $\bar{N}_{ph}$ the total mean number of photo-electron contributing to the coherent flux estimation, we obtain:

$$S_{\partial[X]} = \frac{\partial V_X}{\partial[X]} / V_X + \frac{\partial \bar{N}_{ph}}{\partial[X]} / \bar{N}_{ph} \qquad (3)$$

We now introduce $S_{ph,X}$, the effective sensitivity over the coherent flux. We can express it as the inverse of the SNR over the local coherent flux estimation $\hat{F}_X$:

$$S_{ph,X} = 1/SNR(\hat{F}_X) \qquad (4)$$

Given the mean interferometric intensities shown in Figure 3, it is consistent to assume a photon-noise dominated regime. Under this condition, $SNR(\hat{F}_X)$ is derived from [13]:

$$SNR(\hat{F}_X) \approx V_X\sqrt{\overline{N}_{ph}} \qquad (5)$$

The ratio $S_{ph,X}/S_{\partial[X]}$ gives a good appreciation of the NanoCarb radiometric sensitivity for the given concentration variation $\partial[X]$. Indeed, $(S_{ph,X}/S_{\partial[X]})\partial[X]$ is directly a statistical error over the concentration estimation in the same unit as $\partial[X]$. For example, a ratio of 0.5 for $\partial[CO_2]=1$ ppm means a statistical error over L1b data product equivalent to 0.5 ppm.

Considering Eq. 3 and 5, we obtain:

$$S_{ph,X}/S_{\partial[X]} = \frac{1}{\sqrt{\overline{N}_{ph}}\frac{\partial V_X}{\partial[X]} + \frac{V_X}{\sqrt{\overline{N}_{ph}}}\frac{\partial \overline{N}_{ph}}{\partial[X]}} \qquad (6)$$

We can observe two terms at the denominator of this last expression:

- The left-term shows as expected a statistical error improvement with the total number of photo-electrons as a function of the photon noise. Moreover, this term highlights the evolution of the NanoCarb sensitivity as a function of the contrast variation over the targeted interferometric regions. It requires a maximization of the interferometric Jacobian contrast e.g. in the 5.8 mm OPD region for the $CO_2$ (Figure 3). Only the spectral filtering performed by the interferential filter permits such an optimization.

- The right-term is a second order ($\sim 10^{-3}$ compare to $\sim 10^1$ for the left-term in the worst case). Nevertheless, it illustrates a potential downgrading of the sensitivity when an optimization of the Jacobian contrast decreases the total number of electrons in the targeted region. Thus, a joint trade-off must be carefully achieved to not degrade the SNR when optimizing the spectral bandwidth. In the next subsection, we will neglect the right-term and write:

$$S_{ph,X}/S_{\partial[X]} \approx \frac{1}{\sqrt{\overline{N}_{ph}}\frac{\partial V_X}{\partial[X]}} \qquad (7)$$

On the maximum sensitivity interferometric area of the $CO_2$ band, the expected $\partial V_{XCO2}/(\partial[CO_2] = 1\ ppm)$ is ranged around $10^{-4}$, calling for $\sim 10^8$ electron per ifov for a target random error of 1 ppm. For $\partial[CO_2] = 0.1\ ppm$, $\sim 10^{10}$ electron per ifov are required. We will see in the following sub-section how the NanoCarb concept fulfills this huge number of required photons.

In conclusion, this expression does not show any impact of the instrumental or geophysical biases, and does not allow consequently to accurately choose the interferometric samples and develop a mitigation strategy. That will be investigated in the next section. Nevertheless, it is relevant to derive a first design of the NanoCarb concept by optimizing some main key elements: the spectral bands, and the allocation of pixels between interferometric and spatial samples as we will explain in the next sub-sections.

### 3.2 Level data product total mean number of photo-electrons

Let us now detail the number of photo-electrons at each data product level as a function of the instrumental parameters.

*L0 data product:*

$\overline{N}_{snap}$ the mean number of photo-electrons per frame and per pixel on the FPA (see Eq. 10) is derived from an integration over the spectral bandwidth of the terrestrial radiance (see Figure 2) over the observation solid angle, considering also the spectral transmission of the FP array in the field, the transmission of the front optics, detector characteristics and the exposure time. The interferograms presented in Figure 3 are based on a similar calculation for a terrestrial albedo of 0.2 and show a realistic intensity per pixel in a snapshot acquisition.

*L1a data product:*

$\overline{N}_{snap} n_{FP}$ gives the total mean number of photo-electrons per frame for a given ifov in a single snapshot acquisition. $n_{FP}$ is the number of FP in the array, and thus the number of thumbnails.

*L1b data product:*

As explained previously, the L1b data product is the co-registration of snapshot acquisitions while the ifov shifts across the thumbnail during the orbit. Given $n_{exp}$ the number of co-registered exposures for a given ifov, we can express the total number of photo-electrons $\bar{N}_{ph}$ in L1b data products as:

$$\bar{N}_{ph} = \bar{N}_{snap} n_{FP} n_{exp}; \; [e/ifov] \tag{8}$$

$n_{exp}$ depends on both the number of pixels per thumbnail side $N_S$ and on acquisition framerate. To permit signal co-registration, the minimum framerate must be set to achieve an orbital brooming of one pixel between two exposures. Thus, the maximum number of co-registered exposures for one single ifov is:

$$max\{n_{exp}\} = N_S \times broom_{pix} \tag{9}$$

With $broom_{pix}$ the orbital brooming expressed in pixel.

### 3.3 Estimation of the statistical error over $CO_2$ and $CH_4$ measurements

We based our radiometric computation over the state of the art MCT SWIR FPA 1k 1k NGP [14], which has already been chosen for the MicroCarb mission [15]. Relevant FPA characteristics are presented in Table 2. The FPA is nominally divided into four areas each allocated to a spectral band (typically 512 by 512 pixels). Nevertheless, the most extreme setups investigate one FPA per spectral band, as well as some interesting features of next generation 2k 2k NGP-like [16].

In the nominal configuration, a choice of 64x64 pixels per thumbnail provides a good trade-off between swath and number of interferometric samples, for a 960 μm FP and micro-lens pitch. We achieve 128 or 192 km swath, respectively, for 2 or 3 km of resolution at ground, as well as an 8 by 8 Fabry-Perot interferometer array.

We target an effective acquisition framerate of half a pixel brooming, increasing interpolation capability to balance orbital drift or pointing biases of the platform. We derive the maximum exposure time from this framerate. Table 3 summarizes these elements of design we consider for the radiometric study, in the so-called "nominal" configuration.

We plot in Figure 7 the statistical error on $CO_2$ or $CH_4$ concentration measurement as expressed in Eq. 7, for the configurations summarized in Table 2 and 3, as a function of the percentage of maximum co-registered snapshot acquisitions $max\{n_{exp}\}$ (Eq. 9). We investigate four observation scenarios: 2 and 3 km of ground spatial resolution, and a terrestrial albedo of 0.05 and 0.2.

Table 2 FPA characteristics used in NanoCarb radiometric study, based on NGP features [14]

| FPA format | 1024 × 1024 |
|---|---|
| QE | 0.9 |
| Sensitivity range | 0.5 – 2.5 μm |
| Pixel pitch | 15 μm |
| Readout noise | 170 e- |
| Operating temperature | 170 K |
| Saturation level | 590 ke- |

Table 3 Nominal quarter-NGP NanoCarb band configuration for performance estimation.

| Altitude | 600 km |
|---|---|

| Framerate | 1/2-pixel brooming | |
|---|---|---|
| Dedicated FPA size | $512 \times 512$ | |
| Pixel per thumbnail side $N_S$ | 64 | |
| Number of thumbnail $n_{FP}$ | 64 | |
| **Resolution at ground (ifov)** | **2 km** | **3 km** |
| Swath (fov) | 128 km | 192 km |
| Maximum exposure time | 144.46 ms | 216.69 ms |

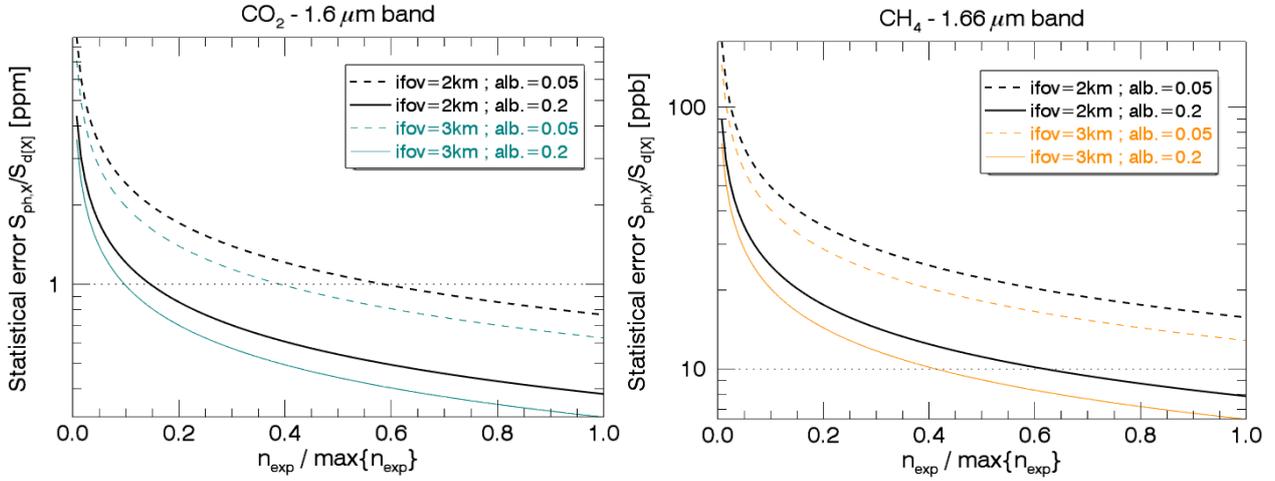

Figure 7 Evolution of the NanoCarb statistical error on $CO_2$ and $CH_4$ measurements as a function of the number of co-registered exposures. On x-axis, 1 means the use of the total number of available frames when the ifov shifts across the thumbnail, while 0 is the snapshot mode.

*Statistical error on $CO_2$ measurement:*

The exploitation of all the available snapshot acquisitions largely fulfills sub-ppm statistical error on $CO_2$ measurement, in all the considered configurations and observation conditions. The performances are even below 0.5 ppm in the more favorable observation conditions (albedo>0.2). Hence, the use of only 50% of the available snapshot acquisitions allows again to reach a sub-ppm statistical error. This is a benefit to mitigate interpolation issues between the different frames, caused for example by platform dis-pointing or jitter.

*Statistical error on CH4 measurement:*

The 10-ppb sensitivity target is only reached in the most favorable observation conditions, with the co-registration of at least 60% of the available frames. We can expect a statistical error around 15 ppb in the worst observation conditions (albedo=0.05), only by co-registering all the available frames.

In conclusion:

- 0.5-1 ppm target for $CO_2$ in statistical error is reached with a nominal allocation of a quarter NGP for the 1.6 μm band.
- The measurement is more complicated in this configuration for $CH_4$, and a target of 10 ppb is only reached in the most favorable cases.

To go further, a very interesting feature of the NanoCarb concept is the dependence of the statistical error to the number of pixels used on the considered spectral band (see Eq. 7, 8 and 9): the number of thumbnails (or FP), $n_{FP}$, can be easily increased by enlarging the FPA area dedicated to the spectral band, any other parameters fixed. Consequently, $\bar{N}_{ph}$ will increase linearly with the size of the dedicated FPA area. Figure 8 shows the statistical error evolution as a function of the number of dedicated pixels, in the case we co-register all the available frames.

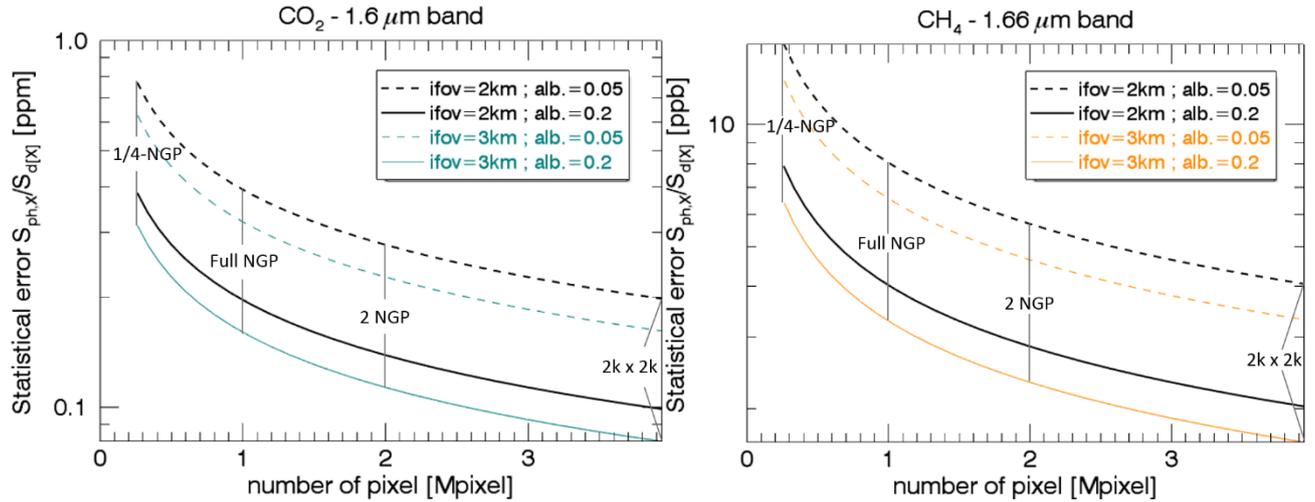

Figure 8 Evolution of the statistical error on $CO_2$ and $CH_4$ measurements as a function of the total number of pixels over the FPA. The statistical error is given for a co-registration of all the available frames $max\{n_{exp}\}$. The number of pixels per thumbnail and the pixel pitch are unchanged. Thus, the number of co-registered exposures is kept constant, as well as the exposure time. The x-axis is equivalent to an increase of the number of thumbnails, and thus, to the number of interferometric samples.

The allocation of an half NGP for the 1.6 µm $CO_2$ band allows to reach a sub-0.5-ppm statistical error in any conditions, while another half NGP for the 1.66 µm $CH_4$ band allows to reach a sub-10-ppb statistical error in most of the conditions, in addition of a relaxation of the required number of co-registered snapshot exposures.

Finally, it is very important to notice that this optimization of the NanoCarb radiometric performances is done with a very faint evolution of the instrumental complexity. As the number of pixels increases for the totality of the spectrometer, we are confident to a linear increase of the volume of the optical part of NanoCarb (from the objective to the FPA), driven by the size of the sensitive area on the detector. As an example, if we evaluate a volume of ~250 cm$^3$ with a NGP, the volume with a 2k 2k FPA will be approximately increased by a factor of 4. Weight and full thermal regulation must be treated as such, which is a real advantage compared to dispersive spectrometry technics.

We demonstrated the radiometric performances of the NanoCarb concept for the targeted sensitivity of $CO_2$ and $CH_4$ in the SCARBO project. In next section, we will introduce a preliminary inverse model to evaluate the main bias impacts over the measurement, and develop a mitigation strategy.

## 4. $CO_2$ TOTAL COLUMN RETRIEVAL

After having studied the random error over the measurement, we aim at considering the absolute accuracy over the NanoCarb L2 data product through an inverse simulation model we have developed.

NanoCarb $CO_2$ and $CH_4$ retrieval strategy and performances are an high-end output of the SCARBO project, consequently investigated during the three allocated years. We present in this last section an in-progress related work, focused only on the $CO_2$ total column retrieval at the NanoCarb L2 data product. We expect to assess some preliminary performances, but also to complete the NanoCarb design strategy concerning an optimal use of the partial interferogram sampling strategy. We will present the three following points:

- Noise propagation during the inversion
- Bias impacts over the $CO_2$ measurement, focused only on the total column of water vapor
- Mitigation strategy and feedback over the NanoCarb design

In addition, we expose in the first sub-section the developed model.

## 4.1 Model

The preliminary model we develop is composed of three blocks, and will be likely improved during the next years: 1) NanoCarb instrumental model, 2) atmospheric model and radiative transfer code, and 3) inversion algorithm.

*NanoCarb instrumental model:*

Currently, we use a simple analytical model of the NanoCarb intensity over the focal plane, derived from the approximate expression of the transmission of a Fabry-Perot. Given $N_{snap}(i,j)$ the number of electron/frame for the considered pixel $(i,j)$ over the focal plane:

$$N_{snap}(i,j) = \eta \int_{\Delta\sigma} \frac{(1/hc\sigma)L_{\sigma,ifov}}{1 + M_\sigma \sin^2\left(\frac{\varphi(i,j)}{2}\right)} d\sigma, \quad [e/frame/pixel] \quad (10)$$

With $\Delta\sigma$ the spectral band, $L_\sigma$ the spectral terrestrial radiance for an ifov, $M_\sigma$ the spectral term of finesse of the Fabry-Perot, and $\eta$ the σ-independent radiometric term, gathering for example exposure time, solid angle, etc. $\varphi(i,j)$ is the phase of the considered pixel. Its OPD is:

$$\delta(i,j) = 2n\varepsilon \cos\theta_{ij} \quad (11)$$

This expression shows the dependence of the OPD for the considered pixel on the thickness of the associated FP of the array, and on the refracted angle for the associated ifov in the total field. Thus, Eq. 10 and 11 allow to generate full images of NanoCarb in the focal plane as presented in Figure 4, as well as a L1b data product model for a single ifov (Figure 6) that will be considered in this section.

In this model, we neglect the Point Spread Function of the micro-lens array, assuming a spatial sub-sampling in the focal plane, and any optical aberrations in the field. We will investigate this point in the next months of the project.

*Atmospheric model and radiative transfer code*

We use the Standard US Model [17] for temperature and pressure profile as well as an atmospheric composition and concentration profile. This model is implemented in the radiative transfer code LBLRTM [18], using the spectroscopic database HITRAN [19]. We simulate NADIR terrestrial radiances with LBLRTM in the 1.6 μm band, taking into account fine spectroscopic effects such as line blending.

For this preliminary work, we assume a clear sky. $CO_2$ retrieval with input measurements of a joint aerosol-dedicated instrument (SPEX [20], developed by SRON) will be implemented later in the framework of SCARBO. By focusing on the 1.6 μm band, we assume also that the atmospheric pressure is perfectly measured in the NanoCarb dedicated band.

*Inversion algorithm*

We use a simple least mean square method to retrieve $CO_2$ concentration, based on a Levenberg-Marquardt algorithm. Its particularity is to work in discrete regions of the Fourier domain, on the contrary to classical inversion strategies considering the radiance space. Thus, we directly retrieve L2 data product of NanoCarb from L1b partial interferograms. The development of an optimized retrieval algorithm is scheduled in the SCARBO project, and will be consequently investigated more precisely in the next years.

## 4.2 Noise propagation

We consider only the favorable observation scenario with a terrestrial albedo of 0.2. The ground spatial resolution of NanoCarb is 2 km. The solar angle is set to 20°. The thickness distribution of the NanoCarb FP array is chosen to sample the region around 5.8 mm where the sensitivity to $CO_2$ is optimal. We assume a perfect L1b data model to retrieve [$CO_2$], meaning that the same model is used both to simulate data and in the retrieval algorithm. We simulate the readout noise with a Gaussian distribution, in addition to a Poisson noise. We perform several retrieval runs in these conditions with a different occurrence of noise for each. The data value of [$CO_2$] is 400 ppm, while the guess value of the retrieval is less than 5%.

Figure 9 shows the interferometric model fitted to the data for a single run, and an histogram of the residuals. As expected the residuals follow a Poisson distribution. Due to the time computation of a single run, we cannot currently converge in a reasonable delay to an accurate value of the propagated noise during the retrieval. Nevertheless, it seems to be around 0.3 ppm for $CO_2$ estimation, that is quite consistent with the radiometric performances of the previous section.

In conclusion, we are confident that the final statistical errors over $[CO_2]$ and $[CH_4]$ measurements are very close to the results of the previous section, without any important noise amplification during the retrieval.

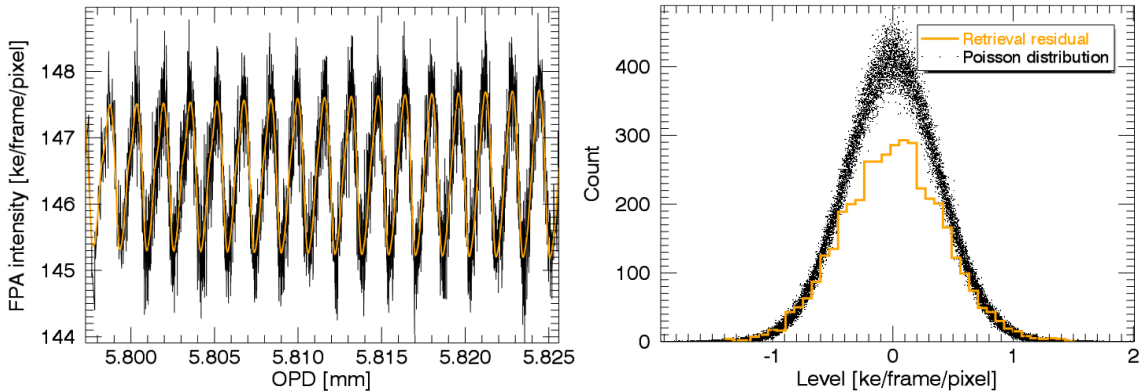

Figure 9 Left: fitted L1b model (orange) to the simulated noisy data for a single retrieval run. Right: histogram of the retrieval residual (orange), compared to a theoretical Poisson distribution.

### 4.3 Impact of the total column of water vapor

Here, we quantify the impact of a misestimate of the total column of $H_2O$ over the $[CO_2]$ retrieval. As we explained, the water vapor is one of the main bias induced over $[CO_2]$ measurement in the 1.6 µm band, assuming a clear sky and a perfect joint measurement of atmospheric pressure. The goal is to derive an estimate of the required accuracy we have to reach.

We target the 5.8 mm region of the interferogram on the 1.6 µm $CO_2$ band. Apart from the bias on $H_2O$, the retrieval L1b model is perfect, and any noises are added to the simulated data. The $[CO_2]$ guess for the retrieval is about 95% of the data value (400 ppm). Figure 10 presents the evolution of the absolute accuracy over $CO_2$ estimate with respect to the bias on water. We notice a strong impact of a water misestimation, because 10% error on water induces a bias greater than 1 ppm over the $CO_2$, while 1% leads to a bias greater than 0.1 ppm.

This observation was well expected if we consider the $CO_2$ interferogram as in Figure 3-left: the Jacobian for 1 ppm of $[CO_2]$ is comparable in flux to the one for 10% of water in the 5.8 mm region. Therefore, both $H_2O$ and $CO_2$ have a significant and comparable impact in these proportions over the fringe intensity. Nevertheless, the contrast of the fringes at 5.8 mm in the water Jacobian is lesser than $CO_2$. Consequently, the water measurement in this interferometric region is potentially less accurate.

This result justifies the need to refine the NanoCarb design to mitigate the water impact, jointly to the development of an adapted retrieval strategy. Especially, we will consider in the next sub-section the selection of several dedicated interferometric regions.

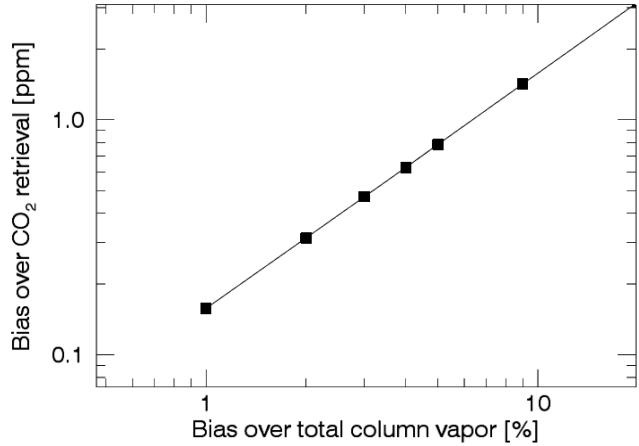

Figure 10 Evolution of the bias over total column of $CO_2$ retrieval as a function of a misestimate of the total column of $H_2O$. The 5.8 mm maximum $CO_2$ sensitivity region of the interferogram is targeted.

**4.4 Bias mitigation strategy**

We investigate in this last sub-section the potential of a partial interferometric sampling to target useful information and drastically reduce bias impacts, on the example of water. In line with this, on top of the 5.8 mm $CO_2$-sensitive region we target the 1.15 mm $H_2O$-sensitive region of the interferogram, as illustrated in Figure 3. This choice is only driven by comparing the fringe contrast over the Jacobian respectively for $CO_2$ and $H_2O$.

Within the nominal configuration of NanoCarb, 8 Fabry-Perot (among 64) are allocated to the sampling of the $H_2O$ region, and the remaining part is dedicated to $CO_2$. This configuration is illustrated in Figure 4, as well as a quite similar L1b interferogram as shown in Figure 6. Currently, this choice is purely arbitrary. We aim at developing an adapted metric to formally optimize the OPD distribution over the different regions in a next step.

For this retrieval run, we consider a perfect inverse model, noiseless data, and a guess concentration of 90% of the real value for the total column of $CO_2$ and $H_2O$. The iterative convergence of the retrieval is presented in Figure 11.

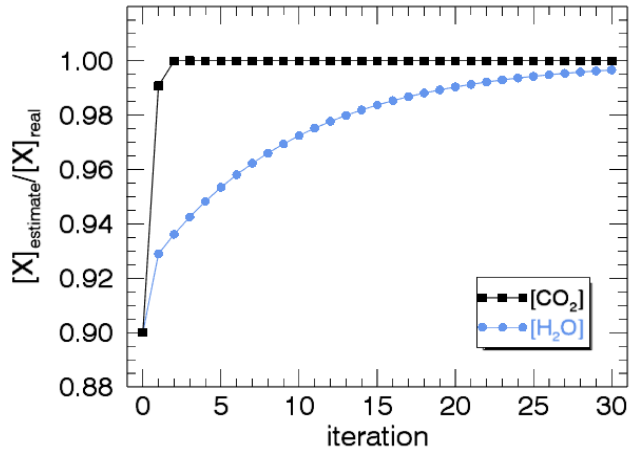

Figure 11 Retrieval convergence of a joint estimation of $[CO_2]$ and $[H_2O]$ on the 1.6 μm band, by sampling two dedicated regions on the interferogram.

Despite a slow convergence of the water retrieval, this very interesting result shows a quick convergence of the $[CO_2]$ retrieval towards a numerically-limited fraction of the real value. It seems that the chosen partial sampling of the interferogram permits in this case a robust degeneracy removing over the $[CO_2]$ retrieval. Furthermore, we achieve in this run a sub-% accuracy over $H_2O$ retrieval.

Finally, we illustrated here the design strategy to optimize the absolute accuracy of NanoCarb. The obtained performances for water mitigation are promising for $CO_2$ measurement in the framework of SCARBO. Both convergence speed and accuracy over $H_2O$ retrieval may be increased, by considering "per piece" retrieval over each dedicated part of the interferogram. Nevertheless, the reached accuracy for water in the 1.6 µm band is positive for [$CH_4$] retrieval. Indeed, the water bias induced in the 1.66 µm band is potentially more important as shown in Figure 3. A joint water retrieval could be beneficial to reach the required accuracy for $CH_4$.

As a conclusion, the presented section is a first step on the performances assessment of NanoCarb. We have illustrated both the design and the retrieval strategy we must consider to optimize such an innovative instrument. In the framework of SCARBO, we are extending and refining this study, by considering a greater number of bias sources and relevant mitigation strategy, in the totality of the spectral bands of NanoCarb (including $O_2$ band, strong $CO_2$ band, and $CH_4$).

First about the atmospheric model improvements are foreseen for:

- Terrestrial surface pressure bias impact
- Atmospheric temperature profile bias impact
- Aerosols measurement inputs from SPEX

Then, about instrumental issues:

- Spatial interpolation strategy for snapshot frames co-registration, and bias induced by platform pointing jitter
- Optical quality in the focal plane and in the field
- Refined detector model, including fine effect such as pixel crosstalk, memory effect, etc.

Finally, about calibration issues:

- Underestimated calibration of the FP thickness, requiring lab experimental measurement of the interferometric plates as investigated in this conference [12]
- Fixed Pattern Noise, dead pixels, RTS over the FPA, photometry and temporal drift of the relevant calibrations.

## 5. CONCLUSION

NanoCarb is an imaging spectrometer concept combining the use of innovative interferometric component and an unusual partial interferogram sampling technic. These two combined features are investigated in the SCARBO project to assess the feasibility of a constellation based on miniaturized payloads to monitor the Earth GHG emissions with a daily revisit and a global coverage. With an optimal use of the available pixels, a NanoCarb-based GHG-sensor achieves a random error over the total column of $CO_2$ and $CH_4$ respectively sub-ppm and below 10 ppb. A very simple optimization of these performances is still possible by increasing the number of dedicated pixels in the FPA. Then, the design strategy implied to accurately choose the targeted sampled regions in the Fourier domain to mitigate geophysical bias impacts over the concentration retrieval. Such a work for water shows a quite complete mitigation of this interferant over the [$CO_2$] retrieval, which demonstrates the potential of the use of partially sampled interferograms. Short term works in the SCARBO timescale consist in a generalization of this strategy to a maximum of geophysical interferants, as well as biases induced by instrumental issues.

## ACKNOWLEDGEMENTS

This project has received funding from the European Union's H2020 research and innovation program under grant agreement No 769032.

The authors would like also to specially thank the FOCUS French label of excellence LabEx FOCUS (ANR-11-LABX-0013) for their funding on parts of this work, as well as for their involvements in these challenging developments.